\documentclass[superscriptaddress,twocolumn,longbibliography,aps,prl,amsmath,amssymb,floatfix,reprint]{revtex4-1}
\usepackage[dvipdfmx]{graphicx}
\usepackage{dcolumn}
\usepackage{bm}
\usepackage[utf8]{inputenc}
\usepackage[T1]{fontenc}
\usepackage{mathptmx}
\usepackage{upgreek}
\usepackage{color}

\begin{document}

\title{Finite temperature effects on the structural stability of Si-doped HfO$_{2}$ using first-principles calculations}

\affiliation{Division of Materials Science, Nara Institute of Science and Technology, Ikoma 630-0192, Japan}
\affiliation{Center for Computational Sciences, University of Tsukuba, Tsukuba 305-8577, Japan}
\affiliation{Department of Computational Science and Technology, Research Organization for Information Science and Technology, Tokyo 105-0013, Japan}
\affiliation{S-Technology Development Center, Tokyo Electron Technology Solutions Ltd.}
\affiliation{Advanced Data Planning Department, Tokyo Electron Ltd.}
\affiliation{Faculty of Pure and Applied Science, University of Tsukuba, Tsukuba 305-8573, Japan}

\author{Y. Harashima}
\email{harashima.yosuke@ms.naist.jp}
\affiliation{Division of Materials Science, Nara Institute of Science and Technology, Ikoma 630-0192, Japan}
\affiliation{Center for Computational Sciences, University of Tsukuba, Tsukuba 305-8577, Japan}

\author{H. Koga}
\affiliation{Department of Computational Science and Technology, Research Organization for Information Science and Technology, Tokyo 105-0013, Japan}

\author{Z. Ni}
\affiliation{S-Technology Development Center, Tokyo Electron Technology Solutions Ltd.}

\author{T. Yonehara}
\affiliation{Department of Computational Science and Technology, Research Organization for Information Science and Technology, Tokyo 105-0013, Japan}

\author{M. Katouda}
\affiliation{Department of Computational Science and Technology, Research Organization for Information Science and Technology, Tokyo 105-0013, Japan}

\author{A. Notake}
\affiliation{Advanced Data Planning Department, Tokyo Electron Ltd.}

\author{H. Matsui}
\affiliation{S-Technology Development Center, Tokyo Electron Technology Solutions Ltd.}

\author{T. Moriya}
\affiliation{Advanced Data Planning Department, Tokyo Electron Ltd.}

\author{M. K. Si}
\affiliation{Center for Computational Sciences, University of Tsukuba, Tsukuba 305-8577, Japan}

\author{R. Hasunuma}
\affiliation{Faculty of Pure and Applied Science, University of Tsukuba, Tsukuba 305-8573, Japan}

\author{A. Uedono}
\affiliation{Faculty of Pure and Applied Science, University of Tsukuba, Tsukuba 305-8573, Japan}

\author{Y. Shigeta}
\affiliation{Center for Computational Sciences, University of Tsukuba, Tsukuba 305-8577, Japan}

\date{\today}

\begin{abstract}
  The structural stabilities of the monoclinic and tetragonal phases of Si-doped HfO$_{2}$ at finite temperatures were analyzed
  using a computational scheme to assess the effects of impurity doping.
  The finite temperature effects considered in this work represented lattice vibration and impurity configuration effects.
  The results show that 6\% Si doping stabilizes the tetragonal phase at room temperature, 
  although a higher concentration of Si is required to stabilize the tetragonal phase at zero temperature.
  These data indicate that lattice vibration and impurity configuration effects are important factors determining structural stability at finite temperatures.
\end{abstract}

\maketitle

\section{Introduction}

Hafnium-based materials, typically HfO$_{2}$, have found numerous applications 
in the semiconductor industry as high-k dielectric compounds~\cite{Wilk2001} and 
are also promising ferroelectric and antiferroelectric materials~\cite{Park2015}. 
These compounds were first employed in the 200s, such as during the 45 nm technology node in 2007~\cite{Mistry2007}
and in certain earlier applications in dynamic random-access memory~\cite{Kim2018}. 
The application of high-k materials allowed the continued scaling of various devices 
by suppressing the gate leakage associated with greater physical thicknesses exhibited by conventional SiO$_{2}$ dielectric materials. 
Since then, HfO$_{2}$ and analogues obtained by doping have become crucial components 
of many advanced device structures, ranging from gate-all-around transistors~\cite{Ernst2006} 
to post-silicon 2D channels such as transition metal dichalcogenides~\cite{Li2019,Xia2017}. 
%% Meanwhile, different phases of polymorphic HfO$_{2}$-based materials possess distinct ferroelectric (FE) or 
%% antiferroelectric (AFE) behaviors~\cite{Park2015}, 
%% rendering them strong candidates for FE random access memory (FeRAM), 
%% FE field effect transistors (FeFET), and various energy-related applications~\cite{Ali2020}. 

The exceptional electronic properties of HfO$_{2}$ appear primarily in the tetragonal and orthorhombic phases. 
Notably, the dielectric constant, $\kappa$, of tetragonal HfO$_{2}$ is $\sim 30$
and so is approximately twice that of stable monoclinic HfO$_{2}$~\cite{Tomida2006}. 
However, the tetragonal and orthorhombic phases are not stable at room temperature~\cite{Wang1992,Tomida2006}.
Consequently, to allow the application of these materials in various devices, 
it is important to develop techniques to stabilize the tetragonal phase.
%% Various approaches have been proposed to stabilize the tetragonal phase. 

%% HfO$_{2}$ has a large dielectic constant, and
%% is one of the promissing candidates as a high-k compound used in semiconductor devices~\cite{Wilk2001}.
%% The high-k compounds are used as insulating films for the gate electrodes in metal-oxide-semiconductor field effect transistors.
%% The gates act as capacitors, and current leakage on the gates compromises the device reliability.
%% To prevent from the leakage, the film need to be thick.
%% However, a thicker film is less susceptible to gate voltage, and device performance is sacrificed.
%% To improve device performance while preventing leakage, 
%% it is necessary to use compounds with higher dielectric constants in the film.

%% HfO$_{2}$ has polymorphs and the dielectric constant depends on the phases.
%% The monoclinic structure is known to be the most stable for the pristine HfO$_{2}$ at room temperature~\cite{Wang1992,Tomida2006}, 
%% whereas the dielectric constant in the tetragonal phase is larger than that in the monoclinic phase~\cite{Materlik2015,Park2017}.
%% In the device applications, it is important to realize the tetragonal phase.

Techniques for controlling the phase have been developed through doping of impurities at different concentration.
As an example, a prior study produced a Si-doped HfO$_{2}$ thin film having a ferroelectric orthogonal structure~\cite{Boescke2011a,Boescke2011b}.
Additionally, recent experimental work fabricated heterogeneous Hf$_{1-x}$Zr$_{x}$O$_{2}$ almost entirely composed of a non-monoclinic phase 
using physical vapor deposition with a composition depth profile suggested by a machine learning model 
(that is, using Bayesian optimization)~\cite{Ni2022}.
Fischer {\it et al.} studied doping with the group IV elements Si, C, Ge, Sn, Ti and Ce into bulk HfO$_{2}$~\cite{Fischer2008a,Fischer2008b} 
and found that Si and Ge stabilize the tetragonal structure.
Doping with Si, P, Ge, Al, Y, Ti, Zr, Gd and Sc was also investigated in HfO$_{2}$ by Lee {\it et al.}~\cite{Lee2008}, 
who identified the microscopic mechanisms by which these dopants stabilize the tetragonal structure on the basis of ionic radii.
In other work~\cite{Barabash2017}, possible metastable phases of HfO$_{2}$ were analyzed using first-principles calculations.
As well, a systematic screening of dopants capable of inducing ferroelectric characteristics in HfO$_{2}$ has been performed~\cite{Batra2017}. 

When assessing the viability of synthesizing a particular phase, 
a theoretical analysis of the structural stability at finite temperatures is required.
In addition, the total energies obtained from first-principles calculations based on density functional theory indicate structural stability at zero temperature.
To investigate the stabilization imparted by doping with impurities at finite temperatures, 
it is necessary to consider both lattice vibration effects and impurity configuration effects.
In a previous study~\cite{Fischer2008b}, 
calculations were performed that incorporated the lattice vibrational free energy based on considering phonon dispersion.
Typically, the impurity configuration is assumed to be homogeneous, 
such that the impurity atoms are periodically aligned within the host lattice.
Recently the authors performed a systematic search for elements capable of stabilizing HfO$_{2}$ and ZrO$_{2}$ using first-principles calculations~\cite{Harashima2022}.
In this prior work, several impurity atoms were included in a supercell and 
the total energies for all possible impurity configurations in the supercell were estimated.
In this manner, the most stable configuration at zero temperature was ascertained.
The results indicated that Si substituted at 6 \% was aligned along linearly and thus not homogeneously distributed throughout the host matrix, 
suggesting that the dopant configuration cannot be neglected when calculating energy values.

The present study investigated the finite temperature phase stability of Si-doped HfO$_{2}$, 
based on a computational scheme that simultaneously considered lattice vibration and impurity configuration effects.
On the basis of the results, 
the transformation between the tetragonal and monoclinic phases of HfO$_{2}$ in association with Si doping is discussed.
The data indicate that 6\% Si doping stabilizes the tetragonal structure at room temperature 
while the monoclinic structure is the most stable at zero temperature.

\section{Methods}

\begin{figure}
  \begin{center}
  \includegraphics[width=\linewidth]{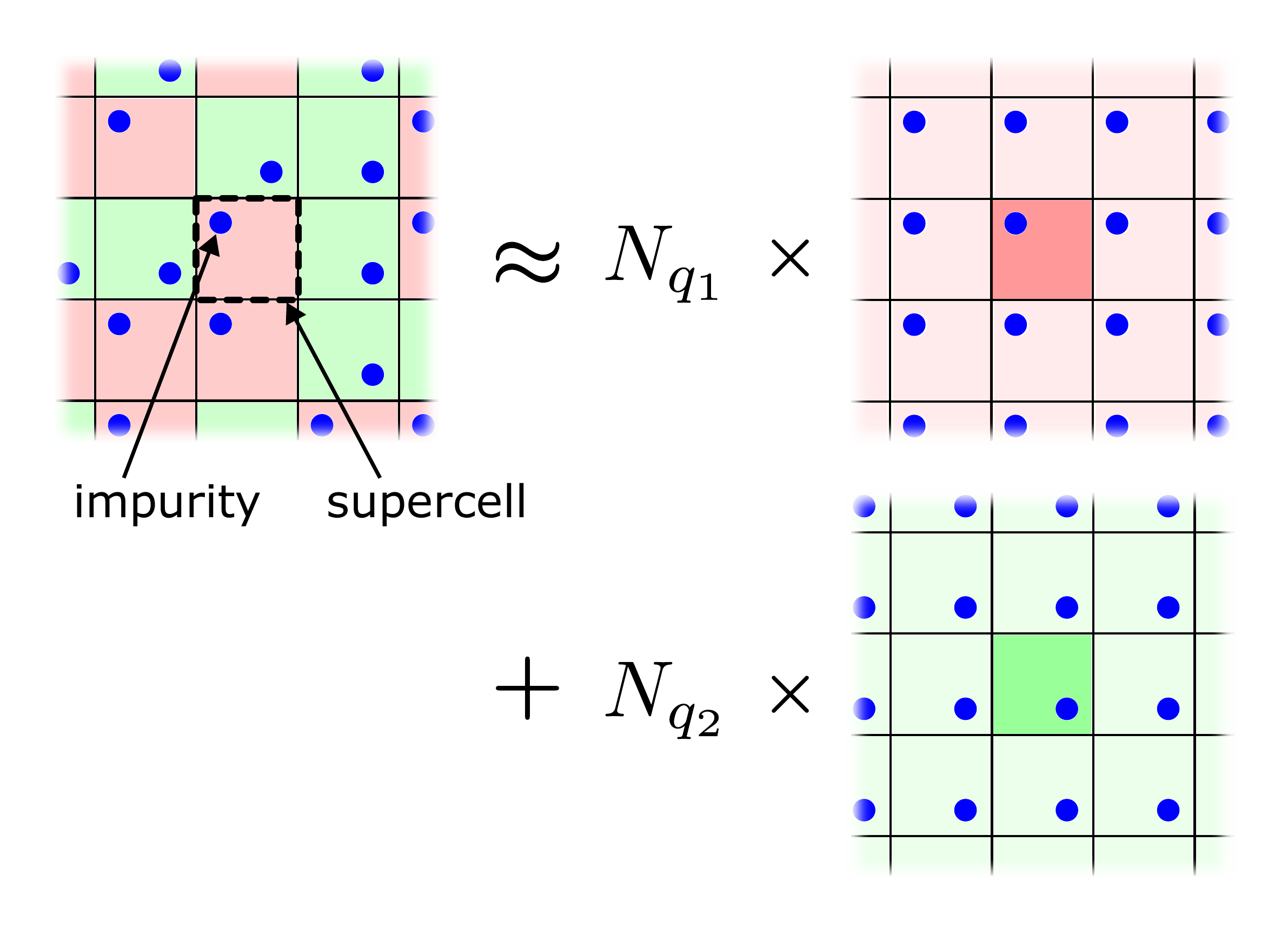}
  \caption{Schematic of the model used to approximate impurity configurations, 
    showing a configuration $\Lambda$ is shown.
    The left-hand side of the figure presents two types of supercells, $q_{1}$ (red) and $q_{2}$ (green), 
    having different impurity configurations. 
    The impurities are indicated by blue spheres.
    The overall system is modelded as a combination of supercells and 
    the contribution from each cell is approximated by that of a piece of a system in which the pieces are aligned periodically.
    Two periodically aligned systems ($q_{1}$-type and $q_{2}$-type) are considered, 
    as shown in the right-hand side of the figure, and 
    the electronic states and vibrational modes for these two systems are calculated.
    Here, $N_{q_{1}}$ and $N_{q_{2}}$ are the numbers of the two types of cells in the entire system.}
  \label{fig:impurity_configuration}
  \end{center}
\end{figure}

This work determined the free energies of impurity substituted systems considering impurity configurations and lattice vibrations.
The calculations are based on a system having impurity configuration $\Lambda$ and lattice vibrational degrees of freedom $\Gamma$.
The state of $\Gamma$ is dependent on $\Lambda$ but this relationship is not explicitly written. 
In this process, the partition function $Z$ is expressed as the sum over all possible of $\Lambda$, 
written as 
\begin{align}
  Z &= \sum_{\Lambda}\sum_{\Gamma}\exp(-\beta E_{\Lambda}(\Gamma))
  \label{eq:partitionfunction_1}
  \\
  &= \sum_{\Lambda} Z_{\Lambda},
  \label{eq:partitionfunction_2}
  \\
  Z_{\Lambda} &\equiv \sum_{\Gamma}\exp(-\beta E_{\Lambda}(\Gamma)).
  \label{eq:partitionfunction_vibration}
\end{align}
Here, $E_{\Lambda}(\Gamma)$ denotes the electronic and vibrational energy of the configuration $\Lambda$ and 
$Z_{\Lambda}$ is the partition function for this configuration.
It is not feasible to calculate the exact value of $Z$ or even of $Z_{\Lambda}$ because the system size is very large (or even infinite at the thermodynamic limit).
Therefore, the system is separated into supercells and the energy of the overall system is approximated by summing over these supercells.
This process can be expressed in terms of $Z_{\Lambda}$ as 
\begin{align}
  Z_{\Lambda} &= \prod_{I} z_{I}.
  \label{eq:partitionfunction_vibration_approximation}
\end{align}
In this case, the impurities are distributed in each supercell $I$.
The energies for a single supercell $I$ can be approximated as the energies for a system of periodically aligned impurities (see Fig.~\ref{fig:impurity_configuration}).
Considering the free energy $f_{I}$ of a periodically aligned supercell, 
we can write
\begin{align}
  Z_{\Lambda} &\approx \prod_{I}\sum_{\gamma_{I}}\exp(-\beta e_{I}(\gamma_{I}))
  \label{eq:partitionfunction_3}
  \\
  &= \prod_{I}\exp(-\beta f_{I})
  \label{eq:partitionfunction_4}
  \\
  &= \exp\left(-\beta\sum_{I}f_{I}\right).
  \label{eq:partitionfunction_5}
\end{align}
Here, $e_{I}$ denotes the energy of supercell $I$.
These energies include contributions from vibrational modes $\gamma_{I}$ in the approximated periodic system, 
such that the free energy $f_{I}$ in cell $I$ is 
\begin{equation}
  f_{I} \equiv -\dfrac{1}{\beta} \ln \sum_{\gamma_{I}}\exp(-\beta e_{I}(\gamma_{I})).
\end{equation}
Note that this $f_{I}$ is the value per supercell.
Because the supercell size is finite, there is only a finite number of impurity configuration patterns that need to be considered.
The partition function for the overall system is therefore written in terms of the free energy $f_{q}$ of each pattern $q$ as
\begin{align}
  Z &\approx \sum_{\Lambda} \exp\left(-\beta\sum_{I}f_{I}\right)
  \label{eq:partitionfunction_6}
  \\
  &= \sum_{\Lambda} \exp\left(-\beta N \sum_{q}n_{q}f_{q}\right)
  \label{eq:partitionfunction_7}
  \\
  &= \sum_{\{n_{q}\}} W_{\{n_{q}\}}\exp\left(-\beta N \sum_{q}n_{q}f_{q}\right),
  \label{eq:partitionfunction_8}
\end{align}
where $n_{q}$ is the proportion of cells having pattern $q$ in the entire system and 
$N$ is the number of supercells which, as noted, will be huge or even infinite at the thermodynamic limit.
The energy value is approximated by summing over the patterns and 
the summation over $\Lambda$ can be obtained by summation over $\{n_{q}\}$, 
as when going from Eq.~\eqref{eq:partitionfunction_7} to \eqref{eq:partitionfunction_8}.
Using Stirling's approximation, 
the weight $W_{\{n_{q}\}}$ can be estimated as
\begin{equation}
  W_{\{n_{q}\}} = \dfrac{N!}{\prod_{q}N_{q}!} \approx \exp\left(-N\sum_{q}n_{q}\ln n_{q}\right).
  \label{eq:weight}
\end{equation}
$N_{q}$ denotes $N n_{q}$.
Substituting Eq.~\eqref{eq:weight} to Eq.~\eqref{eq:partitionfunction_8}, we obtain
\begin{align}
  Z \approx \sum_{\{n_{q}\}} \exp\left(-\beta N \sum_{q}n_{q}\left(f_{q}+\dfrac{1}{\beta}\ln n_{q}\right)\right).
  \label{eq:partitionfunction_9}
\end{align}
At the thermodynamic limit ($N \rightarrow \infty$), 
the most probable set $\{n_{q}\}$ is realized and 
this set satisfies the conditions 
\begin{align}
  &\dfrac{\partial}{\partial n_{q}} \sum_{q} n_{q} \left(f_{q}+\dfrac{1}{\beta}\ln n_{q}\right) = 0,
  \\
  &\sum_{q}n_{q} = 1.
\end{align}
By solving this equation, we obtain 
\begin{equation}
  n_{q} = \dfrac{\exp(-\beta f_{q})}{\sum_{q'}\exp(-\beta f_{q'})}.
  \label{eq:patternratio}
\end{equation}
The dominant contribution in Eq.~\eqref{eq:partitionfunction_9} is given by the set of Eq.~\eqref{eq:patternratio}. 
Taking the logarithm, 
the free energy per supercell can be written as 
\begin{equation}
  f \equiv \dfrac{F}{N} = -\dfrac{1}{\beta N} \ln Z \approx -\dfrac{1}{\beta}\ln\left[\sum_{q}\exp(-\beta f_{q})\right].
  \label{eq:freeenergy_impurity}
\end{equation}
This formula includes the contributions of configurational entropy, 
by summing over $q$, 
and of vibrational free energy via $f_{q}$.
The derivation of this formula is similar to that of the canonical ensemble but uses $f_{q}$, instead of energy.

\section{Results and Discussion}

%% \begin{figure}
%%   \begin{center}
%%   \includegraphics[width=\linewidth]{crystalstructure_hfo2.pdf}
%%   \caption{Monoclinic and tetragonal structures of HfO$_{2}$.}
%%   \label{fig:crystalstructure_hfo2}
%%   \end{center}
%% \end{figure}

The relationships derived in the preceding section were employed to analyze the phase stability of Si-doped HfO$_{2}$.
HfO$_{2}$ can exist as a monoclinic phase having a low dielectric constant 
or a tetragonal phase having a high dielectric constant.
The former is stable at room temperature and 
the transformation from the monoclinic to the tetragonal phases occurs at 2052 K~\cite{Wang2006}.
Here, the free energy values for both phases at finite temperatures were compared.

The $f_{q}$ term was obtained using first-principles calculations, 
employing the Vienna Ab initio Simulation Package (VASP) based on the projector augmented-wave method~\cite{Bloechl1994,Kresse1996,Kresse1999}. 
The generalized gradient approximation was also employed for the exchange-correlation energy functional~\cite{Perdew1996}.
The crystal structures were numerically optimized 
based on preserving the lattice symmetry of the pristine compound and applying 
$a=b$ and $\alpha=\beta=\gamma=\pi/2$ for the tetragonal phase and 
$a=b$ and $\alpha=\gamma=\pi/2$ for the monoclinic phase.
Each dimension of the supercell was twice that of the conventional cell 
($2 \times 2 \times 2$ of the conventional cell), 
which contains 96 atoms in total. 
The impurity configuration patterns could be classified into several groups based on spatial symmetry.
As an example, 2 Hf substitutions in the tetragonal $2 \times 2 \times 2$ supercell can produce ${}_{32}C_{2}=496$ impurity configurations 
that can be classified into nine symmetrically equivalent groups.
The supercell code~\cite{Okhotnikov2016} was used to identify these groups.
The $496$ configurations are separated into 128 patterns $\times$ 2 groups, 64 patterns $\times$ 2 groups, 32 patterns $\times$ 2 groups, and 16 patterns $\times$ 3 groups.
2 Hf substitution into the monoclinic supercell produces 23 groups which contains $32 \times 8$ and $16 \times 15$ groups. 
In the case of practical calculations (that is, for 2 Hf substitutions in the tetragonal structure), 
Eq.~\eqref{eq:freeenergy_impurity} is estimated as
\begin{align}
  f &= -\dfrac{1}{\beta}\ln\left[\sum_{q'=1}^{9}M_{q'} \cdot \exp(-\beta f_{q'})\right],
  \\
  M_{q'} &= \{128, 128, 64, \cdots\},
\end{align}
such that nine $f_{q'}$ values can be calculated using first-principles calculations as explained below.
By comparing $f$ for the tetragonal and monoclinic phases, 
the effect of temperature on phase stability and the structural transformation temperature could be obtained.

In this study, the vibrational free energy of each impurity configuration, $f_{q'}$, was estimated using the Debye-Gr\"{u}neisen model~\cite{Moruzzi1988} based on the equations
\begin{align}
  & F_{q'}(V,T) = E_{\mathrm{el}}(V) 
  \nonumber
  \\
  &\quad - k_{\mathrm{B}}T \left[D(\dfrac{\Theta_{\mathrm{D}}}{T}) 
    - 3\ln \left(1-\exp\left(-\dfrac{\Theta_{\mathrm{D}}}{T}\right)\right)\right] + \dfrac{9}{8}k_{\mathrm{B}}\Theta_{\mathrm{D}},
  \label{eq:freeenergy_debye_groeneisen}
  \\
  & \Theta_{\mathrm{D}} = (6\pi^{2})^{\frac{1}{3}} \dfrac{\hbar}{k_{\mathrm{B}}} v^{\frac{1}{6}} \left(\dfrac{B}{m}\right)^{\frac{1}{2}},
\end{align}
and
\begin{align}
  & D(x) = \dfrac{3}{x^{3}}\int_{0}^{x} dx' \dfrac{x'^{3}}{e^{x'}-1},
\end{align}
where $\Theta_{\mathrm{D}}$ and $D(x)$ are the Debye temperature and Debye function, respectively. 
$v$ and $m$ are averaged volume and mass per atom in supercell, 
and $B$ is a bulk modulus.
The free energy term $F_{q'}$ in Eq.~\eqref{eq:freeenergy_debye_groeneisen} is calculated by first ascertaining the bulk modulus.
This can be accomplished by considering the Murnaghan equation of states
\begin{align}
  & B(p) = B_{0} + B_{0}' p = B_{0}\left(\dfrac{V}{V_{0}}\right)^{-B_{0}'},
  \\
  & E_{\mathrm{el}}(V) = E_{0} - \dfrac{B_{0}}{B_{0}'}V_{0}\left[\dfrac{1}{1-B_{0}'}\left(\dfrac{V}{V_{0}}\right)^{1-B_{0}'} - \dfrac{V}{V_{0}} - \dfrac{B_{0}'}{1-B_{0}'}\right].
\end{align}
The total energies were calculated for volumes expanded and contracted from the stable structure by several percent and 
$B_{0}$ and $B_{0}'$ were obtained by regression of these total energies.
The stable lattice structure for a given temperature was determined as the minimum value of Eq.~\eqref{eq:freeenergy_debye_groeneisen} 
and the free energy values were estimated for an arbitrary temperature.
It should be noted that the present approach involved a systematic error. 
Specifically, the structural transformation temperatures for pristine HfO$_{2}$ and ZrO$_{2}$ were estimated using this process as benchmarks and 
values of 1000 K and 750 K were obtained.
These values are scaled by a factor of 2 with respect to the experimental data, 
and so the calculated transformation temperatures for the doped systems should possible be scaled by a factor of 2 as well.
The accuracy of these calculations could possibly be improved by replacing the process used to assess vibrational modes with first-principles phonon calculations~\cite{Togo2015,Tadano2014,Kuwabara2005,Luo2009}
More sophisticated calculations such as these will be examined in future work.

\begin{figure}
  \begin{center}
  \includegraphics[width=\linewidth]{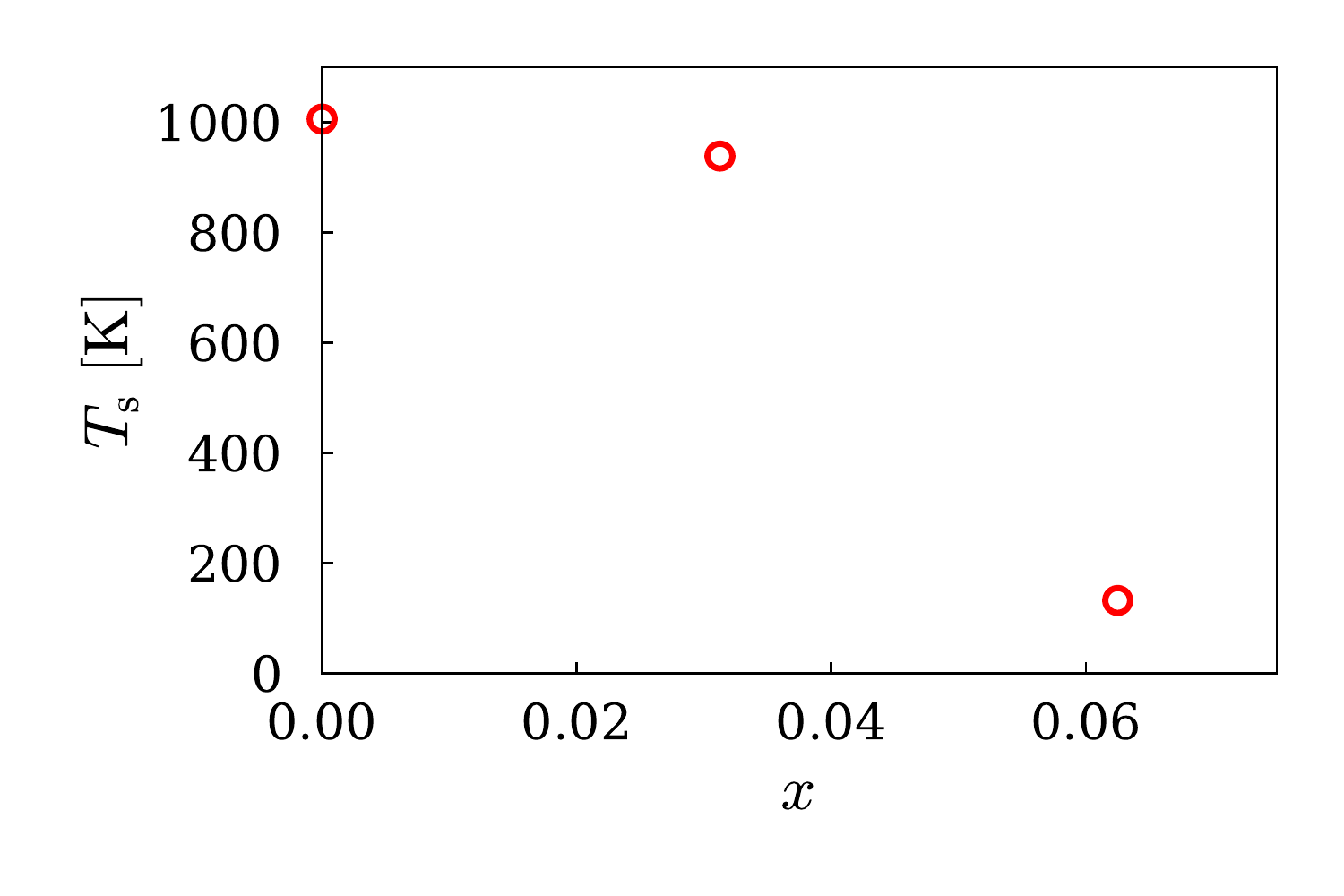}
  \caption{The transformation temperature as a function of the extent of Si doping in Hf$_{1-x}$Si$_{x}$O$_{2}$.}
  \label{fig:ts_impurity_hfo2_si_monoclinic_tetragonal}
  \end{center}
\end{figure}

The transformation temperature was estimated based on the crossing point of the free energies of the tetragonal and monoclinic phases.
Figure~\ref{fig:ts_impurity_hfo2_si_monoclinic_tetragonal} plots the structural transformation temperature $T_{\mathrm{s}}$ of HfO$_{2}$ versus the extent of Si doping and 
demonstrates that the correlation between the two is not linear.
The $T_{\mathrm{s}}$ value for 3\% Si doping is almost the same as that of the pristine compound 
but rapidly decreases as the extent of substitution increases from 3\% to 6\%.
The decreasing trend against Si concentration is consistent with the results of the previous studies.~\cite{Fischer2008b}
The $T_{\mathrm{s}}$ value determined for 6\% substitution was 130 K.
Although, as noted, this temperature may be underestimated and could possibly be scaled by a factor of 2, 
the value would still be lower than room temperature.
This finding indicates that a 6\% Si-doped HfO$_{2}$ dielectric layer in a typical device will exist in the tetragonal phase at a normal operating temperature.

\begin{figure}
  \begin{center}
  \includegraphics[width=\linewidth]{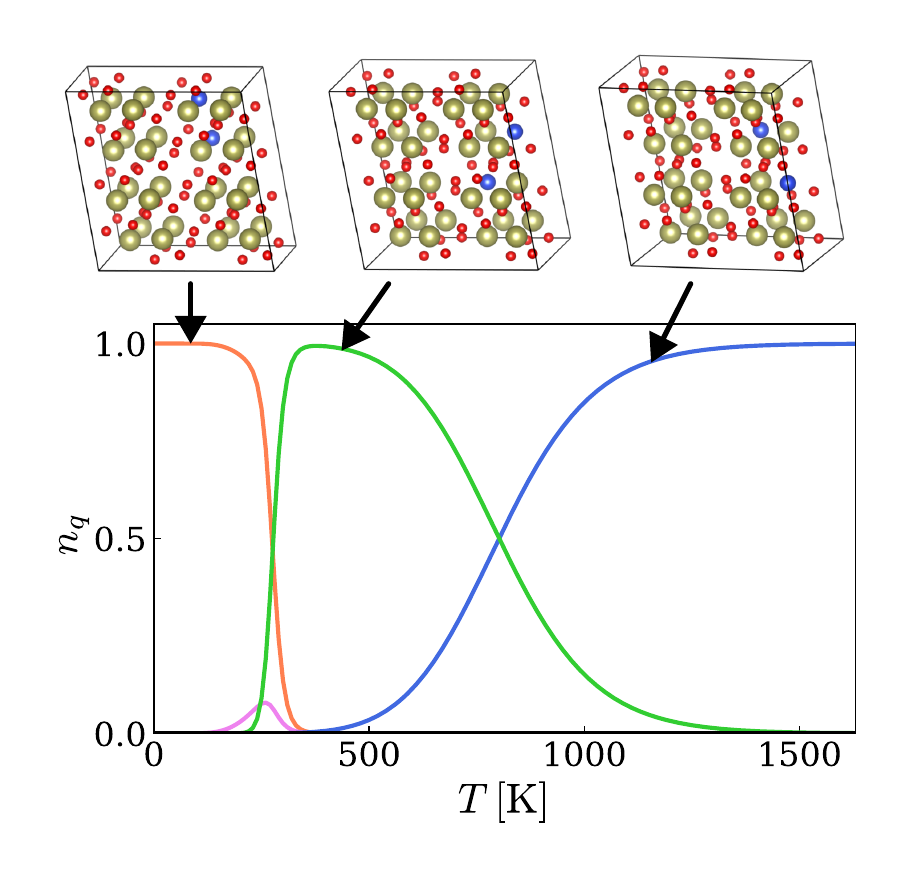}
  \caption{Effect of temperature on the proportion $n_{q}$ of the monoclinic phase (including the multiplicity $M_{q'}$).
    The most probable configurations are shown in the upper panels at the various temperatures. 
    The gold, red and blue spheres denote Hf, O and Si, respectively.}
  \label{fig:concentration_impurity_temperature_hfo2_si_monoclinic}
  \end{center}
\end{figure}

The temperature dependence of the impurity configurations was also examined.
The proportion of each impurity configuration in a system defined as in Eq.~\eqref{eq:patternratio} can be estimated from the $f_{q}$ calculated for each temperature.
In the tetragonal phase, a single configuration (and symmetrically equivalent configurations) was dominant up to $\simeq 1500$ K.
In contrast, in the case of the monoclinic phase, three impurity configurations appeared depending on temperature.
Figure~\ref{fig:concentration_impurity_temperature_hfo2_si_monoclinic} summarizes the effect of temperature on $n_{q}$ for the monoclinic phase.
It is evident that the most stable configuration varied with temperature, 
indicating that the local structure of the impurities affected phase stability.
Even though the calculated transformation temperature was lower than the temperatures at which several configurations were mixed, $\simeq 250$ K or $\simeq 800$ K, 
the temperatures at which such compounds are synthesized will be similar to or much higher than these values. 
Hence, it is important to take into account the appearance of several configurations during the synthetic process.

\section{Conclusion}

This work analyzed the structural stability of Si-doped HfO$_{2}$ 
using the scheme that estimated the free energy and considered impurity configurations and lattice vibrations.
The present process is evidently a useful means of assessing the phase stability of doped polymorphic compounds.
The structural transformation from the monoclinic phase to the tetragonal phase was also investigated based on analyzing the free energy difference.
The transformation temperature values indicate that 
a tetragonal structure with 6\% Si doping will be stable at room temperature 
while a monoclinic structure is more stable at zero temperature.
In the case of a device operating at room temperature, doping with 6\% Si was found to be sufficient to stabilize the tetragonal HfO$_{2}$ phase. 

In general, the physical properties of a compound will be highly correlated with its crystal structure, as demonstrated by the HfO$_{2}$ used in this work as a model case. 
The proposed method in this paper is expected to be applicable to a variety of scenarios.
As an example, in permanent magnets, certain compounds exhibit excellent magnetic properties but are unstable 
and so it is important to stabilize these substances~\cite{Miyake2014,Harashima2016,Harashima2018,Harashima2020}.
In particular, stabilization at the high temperatures at which these compounds are synthesized has recently attracted much attention~\cite{Xing2021}.
We expect that the method demonstrated herein will be effective in such cases 
and could be helpful in the search for structural stabilizers.

\begin{acknowledgments}
This work was supported by JSPS KAKENHI grant number JP22K03449.
The computation was partly conducted using the facilities of the Supercomputer Center at the Institute for Solid State Physics in the University of Tokyo, 
the Multidisciplinary Cooperative Research Program at the Center for Computational Sciences in University of Tsukuba, 
Subsystem B of the ITO system in Kyushu University, 
and Research Center for Computational Science, Okazaki, Japan (Project: 22-IMS-C104).
\end{acknowledgments}

\bibliography{references}

\begin{thebibliography}{34}%
\makeatletter
\providecommand \@ifxundefined [1]{%
 \@ifx{#1\undefined}
}%
\providecommand \@ifnum [1]{%
 \ifnum #1\expandafter \@firstoftwo
 \else \expandafter \@secondoftwo
 \fi
}%
\providecommand \@ifx [1]{%
 \ifx #1\expandafter \@firstoftwo
 \else \expandafter \@secondoftwo
 \fi
}%
\providecommand \natexlab [1]{#1}%
\providecommand \enquote  [1]{``#1''}%
\providecommand \bibnamefont  [1]{#1}%
\providecommand \bibfnamefont [1]{#1}%
\providecommand \citenamefont [1]{#1}%
\providecommand \href@noop [0]{\@secondoftwo}%
\providecommand \href [0]{\begingroup \@sanitize@url \@href}%
\providecommand \@href[1]{\@@startlink{#1}\@@href}%
\providecommand \@@href[1]{\endgroup#1\@@endlink}%
\providecommand \@sanitize@url [0]{\catcode `\\12\catcode `\$12\catcode
  `\&12\catcode `\#12\catcode `\^12\catcode `\_12\catcode `\%12\relax}%
\providecommand \@@startlink[1]{}%
\providecommand \@@endlink[0]{}%
\providecommand \url  [0]{\begingroup\@sanitize@url \@url }%
\providecommand \@url [1]{\endgroup\@href {#1}{\urlprefix }}%
\providecommand \urlprefix  [0]{URL }%
\providecommand \Eprint [0]{\href }%
\providecommand \doibase [0]{http://dx.doi.org/}%
\providecommand \selectlanguage [0]{\@gobble}%
\providecommand \bibinfo  [0]{\@secondoftwo}%
\providecommand \bibfield  [0]{\@secondoftwo}%
\providecommand \translation [1]{[#1]}%
\providecommand \BibitemOpen [0]{}%
\providecommand \bibitemStop [0]{}%
\providecommand \bibitemNoStop [0]{.\EOS\space}%
\providecommand \EOS [0]{\spacefactor3000\relax}%
\providecommand \BibitemShut  [1]{\csname bibitem#1\endcsname}%
\let\auto@bib@innerbib\@empty
%</preamble>
\bibitem [{\citenamefont {Wilk}\ \emph {et~al.}(2001)\citenamefont {Wilk},
  \citenamefont {Wallace},\ and\ \citenamefont {Anthony}}]{Wilk2001}%
  \BibitemOpen
  \bibfield  {author} {\bibinfo {author} {\bibfnamefont {G.~D.}\ \bibnamefont
  {Wilk}}, \bibinfo {author} {\bibfnamefont {R.~M.}\ \bibnamefont {Wallace}}, \
  and\ \bibinfo {author} {\bibfnamefont {J.~M.}\ \bibnamefont {Anthony}},\
  }\href {\doibase 10.1063/1.1361065} {\bibfield  {journal} {\bibinfo
  {journal} {Journal of Applied Physics}\ }\textbf {\bibinfo {volume} {89}},\
  \bibinfo {pages} {5243} (\bibinfo {year} {2001})},\ \Eprint
  {http://arxiv.org/abs/https://doi.org/10.1063/1.1361065}
  {https://doi.org/10.1063/1.1361065} \BibitemShut {NoStop}%
\bibitem [{\citenamefont {Park}\ \emph {et~al.}(2015)\citenamefont {Park},
  \citenamefont {Lee}, \citenamefont {Kim}, \citenamefont {Kim}, \citenamefont
  {Moon}, \citenamefont {Kim}, \citenamefont {Müller}, \citenamefont {Kersch},
  \citenamefont {Schroeder}, \citenamefont {Mikolajick},\ and\ \citenamefont
  {Hwang}}]{Park2015}%
  \BibitemOpen
  \bibfield  {author} {\bibinfo {author} {\bibfnamefont {M.~H.}\ \bibnamefont
  {Park}}, \bibinfo {author} {\bibfnamefont {Y.~H.}\ \bibnamefont {Lee}},
  \bibinfo {author} {\bibfnamefont {H.~J.}\ \bibnamefont {Kim}}, \bibinfo
  {author} {\bibfnamefont {Y.~J.}\ \bibnamefont {Kim}}, \bibinfo {author}
  {\bibfnamefont {T.}~\bibnamefont {Moon}}, \bibinfo {author} {\bibfnamefont
  {K.~D.}\ \bibnamefont {Kim}}, \bibinfo {author} {\bibfnamefont
  {J.}~\bibnamefont {Müller}}, \bibinfo {author} {\bibfnamefont
  {A.}~\bibnamefont {Kersch}}, \bibinfo {author} {\bibfnamefont
  {U.}~\bibnamefont {Schroeder}}, \bibinfo {author} {\bibfnamefont
  {T.}~\bibnamefont {Mikolajick}}, \ and\ \bibinfo {author} {\bibfnamefont
  {C.~S.}\ \bibnamefont {Hwang}},\ }\href {\doibase
  https://doi.org/10.1002/adma.201404531} {\bibfield  {journal} {\bibinfo
  {journal} {Advanced Materials}\ }\textbf {\bibinfo {volume} {27}},\ \bibinfo
  {pages} {1811} (\bibinfo {year} {2015})},\ \Eprint
  {http://arxiv.org/abs/https://onlinelibrary.wiley.com/doi/pdf/10.1002/adma.201404531}
  {https://onlinelibrary.wiley.com/doi/pdf/10.1002/adma.201404531} \BibitemShut
  {NoStop}%
\bibitem [{\citenamefont {Mistry}\ \emph {et~al.}(2007)\citenamefont {Mistry},
  \citenamefont {Allen}, \citenamefont {Auth}, \citenamefont {Beattie},
  \citenamefont {Bergstrom}, \citenamefont {Bost}, \citenamefont {Brazier},
  \citenamefont {Buehler}, \citenamefont {Cappellani}, \citenamefont {Chau},
  \citenamefont {Choi}, \citenamefont {Ding}, \citenamefont {Fischer},
  \citenamefont {Ghani}, \citenamefont {Grover}, \citenamefont {Han},
  \citenamefont {Hanken}, \citenamefont {Hattendorf}, \citenamefont {He},
  \citenamefont {Hicks}, \citenamefont {Huessner}, \citenamefont {Ingerly},
  \citenamefont {Jain}, \citenamefont {James}, \citenamefont {Jong},
  \citenamefont {Joshi}, \citenamefont {Kenyon}, \citenamefont {Kuhn},
  \citenamefont {Lee}, \citenamefont {Liu}, \citenamefont {Maiz}, \citenamefont
  {McIntyre}, \citenamefont {Moon}, \citenamefont {Neirynck}, \citenamefont
  {Pae}, \citenamefont {Parker}, \citenamefont {Parsons}, \citenamefont
  {Prasad}, \citenamefont {Pipes}, \citenamefont {Prince}, \citenamefont
  {Ranade}, \citenamefont {Reynolds}, \citenamefont {Sandford}, \citenamefont
  {Shifren}, \citenamefont {Sebastian}, \citenamefont {Seiple}, \citenamefont
  {Simon}, \citenamefont {Sivakumar}, \citenamefont {Smith}, \citenamefont
  {Thomas}, \citenamefont {Troeger}, \citenamefont {Vandervoorn}, \citenamefont
  {Williams},\ and\ \citenamefont {Zawadzki}}]{Mistry2007}%
  \BibitemOpen
  \bibfield  {author} {\bibinfo {author} {\bibfnamefont {K.}~\bibnamefont
  {Mistry}}, \bibinfo {author} {\bibfnamefont {C.}~\bibnamefont {Allen}},
  \bibinfo {author} {\bibfnamefont {C.}~\bibnamefont {Auth}}, \bibinfo {author}
  {\bibfnamefont {B.}~\bibnamefont {Beattie}}, \bibinfo {author} {\bibfnamefont
  {D.}~\bibnamefont {Bergstrom}}, \bibinfo {author} {\bibfnamefont
  {M.}~\bibnamefont {Bost}}, \bibinfo {author} {\bibfnamefont {M.}~\bibnamefont
  {Brazier}}, \bibinfo {author} {\bibfnamefont {M.}~\bibnamefont {Buehler}},
  \bibinfo {author} {\bibfnamefont {A.}~\bibnamefont {Cappellani}}, \bibinfo
  {author} {\bibfnamefont {R.}~\bibnamefont {Chau}}, \bibinfo {author}
  {\bibfnamefont {C.-H.}\ \bibnamefont {Choi}}, \bibinfo {author}
  {\bibfnamefont {G.}~\bibnamefont {Ding}}, \bibinfo {author} {\bibfnamefont
  {K.}~\bibnamefont {Fischer}}, \bibinfo {author} {\bibfnamefont
  {T.}~\bibnamefont {Ghani}}, \bibinfo {author} {\bibfnamefont
  {R.}~\bibnamefont {Grover}}, \bibinfo {author} {\bibfnamefont
  {W.}~\bibnamefont {Han}}, \bibinfo {author} {\bibfnamefont {D.}~\bibnamefont
  {Hanken}}, \bibinfo {author} {\bibfnamefont {M.}~\bibnamefont {Hattendorf}},
  \bibinfo {author} {\bibfnamefont {J.}~\bibnamefont {He}}, \bibinfo {author}
  {\bibfnamefont {J.}~\bibnamefont {Hicks}}, \bibinfo {author} {\bibfnamefont
  {R.}~\bibnamefont {Huessner}}, \bibinfo {author} {\bibfnamefont
  {D.}~\bibnamefont {Ingerly}}, \bibinfo {author} {\bibfnamefont
  {P.}~\bibnamefont {Jain}}, \bibinfo {author} {\bibfnamefont {R.}~\bibnamefont
  {James}}, \bibinfo {author} {\bibfnamefont {L.}~\bibnamefont {Jong}},
  \bibinfo {author} {\bibfnamefont {S.}~\bibnamefont {Joshi}}, \bibinfo
  {author} {\bibfnamefont {C.}~\bibnamefont {Kenyon}}, \bibinfo {author}
  {\bibfnamefont {K.}~\bibnamefont {Kuhn}}, \bibinfo {author} {\bibfnamefont
  {K.}~\bibnamefont {Lee}}, \bibinfo {author} {\bibfnamefont {H.}~\bibnamefont
  {Liu}}, \bibinfo {author} {\bibfnamefont {J.}~\bibnamefont {Maiz}}, \bibinfo
  {author} {\bibfnamefont {B.}~\bibnamefont {McIntyre}}, \bibinfo {author}
  {\bibfnamefont {P.}~\bibnamefont {Moon}}, \bibinfo {author} {\bibfnamefont
  {J.}~\bibnamefont {Neirynck}}, \bibinfo {author} {\bibfnamefont
  {S.}~\bibnamefont {Pae}}, \bibinfo {author} {\bibfnamefont {C.}~\bibnamefont
  {Parker}}, \bibinfo {author} {\bibfnamefont {D.}~\bibnamefont {Parsons}},
  \bibinfo {author} {\bibfnamefont {C.}~\bibnamefont {Prasad}}, \bibinfo
  {author} {\bibfnamefont {L.}~\bibnamefont {Pipes}}, \bibinfo {author}
  {\bibfnamefont {M.}~\bibnamefont {Prince}}, \bibinfo {author} {\bibfnamefont
  {P.}~\bibnamefont {Ranade}}, \bibinfo {author} {\bibfnamefont
  {T.}~\bibnamefont {Reynolds}}, \bibinfo {author} {\bibfnamefont
  {J.}~\bibnamefont {Sandford}}, \bibinfo {author} {\bibfnamefont
  {L.}~\bibnamefont {Shifren}}, \bibinfo {author} {\bibfnamefont
  {J.}~\bibnamefont {Sebastian}}, \bibinfo {author} {\bibfnamefont
  {J.}~\bibnamefont {Seiple}}, \bibinfo {author} {\bibfnamefont
  {D.}~\bibnamefont {Simon}}, \bibinfo {author} {\bibfnamefont
  {S.}~\bibnamefont {Sivakumar}}, \bibinfo {author} {\bibfnamefont
  {P.}~\bibnamefont {Smith}}, \bibinfo {author} {\bibfnamefont
  {C.}~\bibnamefont {Thomas}}, \bibinfo {author} {\bibfnamefont
  {T.}~\bibnamefont {Troeger}}, \bibinfo {author} {\bibfnamefont
  {P.}~\bibnamefont {Vandervoorn}}, \bibinfo {author} {\bibfnamefont
  {S.}~\bibnamefont {Williams}}, \ and\ \bibinfo {author} {\bibfnamefont
  {K.}~\bibnamefont {Zawadzki}},\ }in\ \href {\doibase
  10.1109/IEDM.2007.4418914} {\emph {\bibinfo {booktitle} {2007 IEEE
  International Electron Devices Meeting}}}\ (\bibinfo {year} {2007})\ pp.\
  \bibinfo {pages} {247--250}\BibitemShut {NoStop}%
\bibitem [{\citenamefont {Kim}\ and\ \citenamefont {Popovici}(2018)}]{Kim2018}%
  \BibitemOpen
  \bibfield  {author} {\bibinfo {author} {\bibfnamefont {S.~K.}\ \bibnamefont
  {Kim}}\ and\ \bibinfo {author} {\bibfnamefont {M.}~\bibnamefont {Popovici}},\
  }\href {\doibase 10.1557/mrs.2018.95} {\bibfield  {journal} {\bibinfo
  {journal} {MRS Bulletin}\ }\textbf {\bibinfo {volume} {43}},\ \bibinfo
  {pages} {334} (\bibinfo {year} {2018})}\BibitemShut {NoStop}%
\bibitem [{\citenamefont {Ernst}\ \emph {et~al.}(2006)\citenamefont {Ernst},
  \citenamefont {Dupre}, \citenamefont {Isheden}, \citenamefont {Bernard},
  \citenamefont {Ritzenthaler}, \citenamefont {Maffini-Alvaro}, \citenamefont
  {Barbe}, \citenamefont {De~Crecy}, \citenamefont {Toffoli}, \citenamefont
  {Vizioz}, \citenamefont {Borel}, \citenamefont {Andrieu}, \citenamefont
  {Delaye}, \citenamefont {Lafond}, \citenamefont {Rabille}, \citenamefont
  {Hartmann}, \citenamefont {Rivoire}, \citenamefont {Guillaumot},
  \citenamefont {Suhm}, \citenamefont {Rivallin}, \citenamefont {Faynot},
  \citenamefont {Ghibaudo},\ and\ \citenamefont {Deleonibus}}]{Ernst2006}%
  \BibitemOpen
  \bibfield  {author} {\bibinfo {author} {\bibfnamefont {T.}~\bibnamefont
  {Ernst}}, \bibinfo {author} {\bibfnamefont {C.}~\bibnamefont {Dupre}},
  \bibinfo {author} {\bibfnamefont {C.}~\bibnamefont {Isheden}}, \bibinfo
  {author} {\bibfnamefont {E.}~\bibnamefont {Bernard}}, \bibinfo {author}
  {\bibfnamefont {R.}~\bibnamefont {Ritzenthaler}}, \bibinfo {author}
  {\bibfnamefont {V.}~\bibnamefont {Maffini-Alvaro}}, \bibinfo {author}
  {\bibfnamefont {J.-C.}\ \bibnamefont {Barbe}}, \bibinfo {author}
  {\bibfnamefont {F.}~\bibnamefont {De~Crecy}}, \bibinfo {author}
  {\bibfnamefont {A.}~\bibnamefont {Toffoli}}, \bibinfo {author} {\bibfnamefont
  {C.}~\bibnamefont {Vizioz}}, \bibinfo {author} {\bibfnamefont
  {S.}~\bibnamefont {Borel}}, \bibinfo {author} {\bibfnamefont
  {F.}~\bibnamefont {Andrieu}}, \bibinfo {author} {\bibfnamefont
  {V.}~\bibnamefont {Delaye}}, \bibinfo {author} {\bibfnamefont
  {D.}~\bibnamefont {Lafond}}, \bibinfo {author} {\bibfnamefont
  {G.}~\bibnamefont {Rabille}}, \bibinfo {author} {\bibfnamefont {J.-M.}\
  \bibnamefont {Hartmann}}, \bibinfo {author} {\bibfnamefont {M.}~\bibnamefont
  {Rivoire}}, \bibinfo {author} {\bibfnamefont {B.}~\bibnamefont {Guillaumot}},
  \bibinfo {author} {\bibfnamefont {A.}~\bibnamefont {Suhm}}, \bibinfo {author}
  {\bibfnamefont {P.}~\bibnamefont {Rivallin}}, \bibinfo {author}
  {\bibfnamefont {O.}~\bibnamefont {Faynot}}, \bibinfo {author} {\bibfnamefont
  {G.}~\bibnamefont {Ghibaudo}}, \ and\ \bibinfo {author} {\bibfnamefont
  {S.}~\bibnamefont {Deleonibus}},\ }in\ \href {\doibase
  10.1109/IEDM.2006.346955} {\emph {\bibinfo {booktitle} {2006 International
  Electron Devices Meeting}}}\ (\bibinfo {year} {2006})\ pp.\ \bibinfo {pages}
  {1--4}\BibitemShut {NoStop}%
\bibitem [{\citenamefont {Li}\ \emph {et~al.}(2019)\citenamefont {Li},
  \citenamefont {Zhou}, \citenamefont {Cai}, \citenamefont {Yu}, \citenamefont
  {Zhang}, \citenamefont {Fang}, \citenamefont {Li}, \citenamefont {Wu},
  \citenamefont {Chen}, \citenamefont {Xie}, \citenamefont {Ma}, \citenamefont
  {Dai}, \citenamefont {Wu}, \citenamefont {Zhao}, \citenamefont {Wang},
  \citenamefont {He}, \citenamefont {Pan}, \citenamefont {Shi}, \citenamefont
  {Wang}, \citenamefont {Chen}, \citenamefont {Nagashio}, \citenamefont
  {Duan},\ and\ \citenamefont {Wang}}]{Li2019}%
  \BibitemOpen
  \bibfield  {author} {\bibinfo {author} {\bibfnamefont {W.}~\bibnamefont
  {Li}}, \bibinfo {author} {\bibfnamefont {J.}~\bibnamefont {Zhou}}, \bibinfo
  {author} {\bibfnamefont {S.}~\bibnamefont {Cai}}, \bibinfo {author}
  {\bibfnamefont {Z.}~\bibnamefont {Yu}}, \bibinfo {author} {\bibfnamefont
  {J.}~\bibnamefont {Zhang}}, \bibinfo {author} {\bibfnamefont
  {N.}~\bibnamefont {Fang}}, \bibinfo {author} {\bibfnamefont {T.}~\bibnamefont
  {Li}}, \bibinfo {author} {\bibfnamefont {Y.}~\bibnamefont {Wu}}, \bibinfo
  {author} {\bibfnamefont {T.}~\bibnamefont {Chen}}, \bibinfo {author}
  {\bibfnamefont {X.}~\bibnamefont {Xie}}, \bibinfo {author} {\bibfnamefont
  {K.}~\bibnamefont {Ma}, \bibfnamefont {Haibo~Yan}}, \bibinfo {author}
  {\bibfnamefont {N.}~\bibnamefont {Dai}}, \bibinfo {author} {\bibfnamefont
  {X.}~\bibnamefont {Wu}}, \bibinfo {author} {\bibfnamefont {H.}~\bibnamefont
  {Zhao}}, \bibinfo {author} {\bibfnamefont {Z.}~\bibnamefont {Wang}}, \bibinfo
  {author} {\bibfnamefont {D.}~\bibnamefont {He}}, \bibinfo {author}
  {\bibfnamefont {L.}~\bibnamefont {Pan}}, \bibinfo {author} {\bibfnamefont
  {Y.}~\bibnamefont {Shi}}, \bibinfo {author} {\bibfnamefont {P.}~\bibnamefont
  {Wang}}, \bibinfo {author} {\bibfnamefont {W.}~\bibnamefont {Chen}}, \bibinfo
  {author} {\bibfnamefont {K.}~\bibnamefont {Nagashio}}, \bibinfo {author}
  {\bibfnamefont {X.}~\bibnamefont {Duan}}, \ and\ \bibinfo {author}
  {\bibfnamefont {X.}~\bibnamefont {Wang}},\ }\href {\doibase
  10.1038/s41928-019-0334-y} {\bibfield  {journal} {\bibinfo  {journal} {Nature
  Electronics}\ }\textbf {\bibinfo {volume} {2}},\ \bibinfo {pages} {563}
  (\bibinfo {year} {2019})}\BibitemShut {NoStop}%
\bibitem [{\citenamefont {Xia}\ \emph {et~al.}(2017)\citenamefont {Xia},
  \citenamefont {Feng}, \citenamefont {Ng}, \citenamefont {Wang}, \citenamefont
  {Chi}, \citenamefont {Li}, \citenamefont {He}, \citenamefont {Liu},\ and\
  \citenamefont {Ang}}]{Xia2017}%
  \BibitemOpen
  \bibfield  {author} {\bibinfo {author} {\bibfnamefont {P.}~\bibnamefont
  {Xia}}, \bibinfo {author} {\bibfnamefont {X.}~\bibnamefont {Feng}}, \bibinfo
  {author} {\bibfnamefont {R.~J.}\ \bibnamefont {Ng}}, \bibinfo {author}
  {\bibfnamefont {S.}~\bibnamefont {Wang}}, \bibinfo {author} {\bibfnamefont
  {D.}~\bibnamefont {Chi}}, \bibinfo {author} {\bibfnamefont {C.}~\bibnamefont
  {Li}}, \bibinfo {author} {\bibfnamefont {Z.}~\bibnamefont {He}}, \bibinfo
  {author} {\bibfnamefont {X.}~\bibnamefont {Liu}}, \ and\ \bibinfo {author}
  {\bibfnamefont {K.-W.}\ \bibnamefont {Ang}},\ }\href {\doibase
  10.1038/srep40669} {\bibfield  {journal} {\bibinfo  {journal} {Scientific
  Reports}\ }\textbf {\bibinfo {volume} {7}},\ \bibinfo {pages} {40669}
  (\bibinfo {year} {2017})}\BibitemShut {NoStop}%
\bibitem [{\citenamefont {Tomida}\ \emph {et~al.}(2006)\citenamefont {Tomida},
  \citenamefont {Kita},\ and\ \citenamefont {Toriumi}}]{Tomida2006}%
  \BibitemOpen
  \bibfield  {author} {\bibinfo {author} {\bibfnamefont {K.}~\bibnamefont
  {Tomida}}, \bibinfo {author} {\bibfnamefont {K.}~\bibnamefont {Kita}}, \ and\
  \bibinfo {author} {\bibfnamefont {A.}~\bibnamefont {Toriumi}},\ }\href
  {\doibase 10.1063/1.2355471} {\bibfield  {journal} {\bibinfo  {journal}
  {Applied Physics Letters}\ }\textbf {\bibinfo {volume} {89}},\ \bibinfo
  {pages} {142902} (\bibinfo {year} {2006})},\ \Eprint
  {http://arxiv.org/abs/https://doi.org/10.1063/1.2355471}
  {https://doi.org/10.1063/1.2355471} \BibitemShut {NoStop}%
\bibitem [{\citenamefont {Wang}\ \emph {et~al.}(1992)\citenamefont {Wang},
  \citenamefont {Li},\ and\ \citenamefont {Stevens}}]{Wang1992}%
  \BibitemOpen
  \bibfield  {author} {\bibinfo {author} {\bibfnamefont {J.}~\bibnamefont
  {Wang}}, \bibinfo {author} {\bibfnamefont {H.~P.}\ \bibnamefont {Li}}, \ and\
  \bibinfo {author} {\bibfnamefont {R.}~\bibnamefont {Stevens}},\ }\href@noop
  {} {\bibfield  {journal} {\bibinfo  {journal} {Journal of Materials Science}\
  }\textbf {\bibinfo {volume} {27}},\ \bibinfo {pages} {5397} (\bibinfo {year}
  {1992})}\BibitemShut {NoStop}%
\bibitem [{\citenamefont {Böscke}\ \emph
  {et~al.}(2011{\natexlab{a}})\citenamefont {Böscke}, \citenamefont {Müller},
  \citenamefont {Bräuhaus}, \citenamefont {Schröder},\ and\ \citenamefont
  {Böttger}}]{Boescke2011a}%
  \BibitemOpen
  \bibfield  {author} {\bibinfo {author} {\bibfnamefont {T.~S.}\ \bibnamefont
  {Böscke}}, \bibinfo {author} {\bibfnamefont {J.}~\bibnamefont {Müller}},
  \bibinfo {author} {\bibfnamefont {D.}~\bibnamefont {Bräuhaus}}, \bibinfo
  {author} {\bibfnamefont {U.}~\bibnamefont {Schröder}}, \ and\ \bibinfo
  {author} {\bibfnamefont {U.}~\bibnamefont {Böttger}},\ }\href {\doibase
  10.1063/1.3634052} {\bibfield  {journal} {\bibinfo  {journal} {Applied
  Physics Letters}\ }\textbf {\bibinfo {volume} {99}},\ \bibinfo {pages}
  {102903} (\bibinfo {year} {2011}{\natexlab{a}})},\ \Eprint
  {http://arxiv.org/abs/https://doi.org/10.1063/1.3634052}
  {https://doi.org/10.1063/1.3634052} \BibitemShut {NoStop}%
\bibitem [{\citenamefont {Böscke}\ \emph
  {et~al.}(2011{\natexlab{b}})\citenamefont {Böscke}, \citenamefont
  {Teichert}, \citenamefont {Bräuhaus}, \citenamefont {Müller}, \citenamefont
  {Schröder}, \citenamefont {Böttger},\ and\ \citenamefont
  {Mikolajick}}]{Boescke2011b}%
  \BibitemOpen
  \bibfield  {author} {\bibinfo {author} {\bibfnamefont {T.~S.}\ \bibnamefont
  {Böscke}}, \bibinfo {author} {\bibfnamefont {S.}~\bibnamefont {Teichert}},
  \bibinfo {author} {\bibfnamefont {D.}~\bibnamefont {Bräuhaus}}, \bibinfo
  {author} {\bibfnamefont {J.}~\bibnamefont {Müller}}, \bibinfo {author}
  {\bibfnamefont {U.}~\bibnamefont {Schröder}}, \bibinfo {author}
  {\bibfnamefont {U.}~\bibnamefont {Böttger}}, \ and\ \bibinfo {author}
  {\bibfnamefont {T.}~\bibnamefont {Mikolajick}},\ }\href {\doibase
  10.1063/1.3636434} {\bibfield  {journal} {\bibinfo  {journal} {Applied
  Physics Letters}\ }\textbf {\bibinfo {volume} {99}},\ \bibinfo {pages}
  {112904} (\bibinfo {year} {2011}{\natexlab{b}})},\ \Eprint
  {http://arxiv.org/abs/https://doi.org/10.1063/1.3636434}
  {https://doi.org/10.1063/1.3636434} \BibitemShut {NoStop}%
\bibitem [{\citenamefont {Ni}\ and\ \citenamefont {Matsui}(2022)}]{Ni2022}%
  \BibitemOpen
  \bibfield  {author} {\bibinfo {author} {\bibfnamefont {Z.}~\bibnamefont
  {Ni}}\ and\ \bibinfo {author} {\bibfnamefont {H.}~\bibnamefont {Matsui}},\
  }\href {\doibase 10.35848/1347-4065/ac64e4} {\bibfield  {journal} {\bibinfo
  {journal} {Japanese Journal of Applied Physics}\ }\textbf {\bibinfo {volume}
  {61}},\ \bibinfo {pages} {SH1009} (\bibinfo {year} {2022})}\BibitemShut
  {NoStop}%
\bibitem [{\citenamefont {Fischer}\ and\ \citenamefont
  {Kersch}(2008{\natexlab{a}})}]{Fischer2008a}%
  \BibitemOpen
  \bibfield  {author} {\bibinfo {author} {\bibfnamefont {D.}~\bibnamefont
  {Fischer}}\ and\ \bibinfo {author} {\bibfnamefont {A.}~\bibnamefont
  {Kersch}},\ }\href {\doibase 10.1063/1.2828696} {\bibfield  {journal}
  {\bibinfo  {journal} {Applied Physics Letters}\ }\textbf {\bibinfo {volume}
  {92}},\ \bibinfo {pages} {012908} (\bibinfo {year} {2008}{\natexlab{a}})},\
  \Eprint
  {http://arxiv.org/abs/https://aip.scitation.org/doi/pdf/10.1063/1.2828696}
  {https://aip.scitation.org/doi/pdf/10.1063/1.2828696} \BibitemShut {NoStop}%
\bibitem [{\citenamefont {Fischer}\ and\ \citenamefont
  {Kersch}(2008{\natexlab{b}})}]{Fischer2008b}%
  \BibitemOpen
  \bibfield  {author} {\bibinfo {author} {\bibfnamefont {D.}~\bibnamefont
  {Fischer}}\ and\ \bibinfo {author} {\bibfnamefont {A.}~\bibnamefont
  {Kersch}},\ }\href {\doibase 10.1063/1.2999352} {\bibfield  {journal}
  {\bibinfo  {journal} {Journal of Applied Physics}\ }\textbf {\bibinfo
  {volume} {104}},\ \bibinfo {pages} {084104} (\bibinfo {year}
  {2008}{\natexlab{b}})},\ \Eprint
  {http://arxiv.org/abs/https://doi.org/10.1063/1.2999352}
  {https://doi.org/10.1063/1.2999352} \BibitemShut {NoStop}%
\bibitem [{\citenamefont {Lee}\ \emph {et~al.}(2008)\citenamefont {Lee},
  \citenamefont {Cho}, \citenamefont {Lee}, \citenamefont {Hwang},\ and\
  \citenamefont {Han}}]{Lee2008}%
  \BibitemOpen
  \bibfield  {author} {\bibinfo {author} {\bibfnamefont {C.-K.}\ \bibnamefont
  {Lee}}, \bibinfo {author} {\bibfnamefont {E.}~\bibnamefont {Cho}}, \bibinfo
  {author} {\bibfnamefont {H.-S.}\ \bibnamefont {Lee}}, \bibinfo {author}
  {\bibfnamefont {C.~S.}\ \bibnamefont {Hwang}}, \ and\ \bibinfo {author}
  {\bibfnamefont {S.}~\bibnamefont {Han}},\ }\href {\doibase
  10.1103/PhysRevB.78.012102} {\bibfield  {journal} {\bibinfo  {journal} {Phys.
  Rev. B}\ }\textbf {\bibinfo {volume} {78}},\ \bibinfo {pages} {012102}
  (\bibinfo {year} {2008})}\BibitemShut {NoStop}%
\bibitem [{\citenamefont {Barabash}(2017)}]{Barabash2017}%
  \BibitemOpen
  \bibfield  {author} {\bibinfo {author} {\bibfnamefont {S.~V.}\ \bibnamefont
  {Barabash}},\ }\href@noop {} {\bibfield  {journal} {\bibinfo  {journal} {J.
  Comput. Electron.}\ }\textbf {\bibinfo {volume} {16}},\ \bibinfo {pages}
  {1227} (\bibinfo {year} {2017})}\BibitemShut {NoStop}%
\bibitem [{\citenamefont {Batra}\ \emph {et~al.}(2017)\citenamefont {Batra},
  \citenamefont {Huan}, \citenamefont {Rossetti},\ and\ \citenamefont
  {Ramprasad}}]{Batra2017}%
  \BibitemOpen
  \bibfield  {author} {\bibinfo {author} {\bibfnamefont {R.}~\bibnamefont
  {Batra}}, \bibinfo {author} {\bibfnamefont {T.~D.}\ \bibnamefont {Huan}},
  \bibinfo {author} {\bibfnamefont {G.~A.~J.}\ \bibnamefont {Rossetti}}, \ and\
  \bibinfo {author} {\bibfnamefont {R.}~\bibnamefont {Ramprasad}},\ }\href
  {\doibase 10.1021/acs.chemmater.7b02835} {\bibfield  {journal} {\bibinfo
  {journal} {Chemistry of Materials}\ }\textbf {\bibinfo {volume} {29}},\
  \bibinfo {pages} {9102} (\bibinfo {year} {2017})},\ \Eprint
  {http://arxiv.org/abs/https://doi.org/10.1021/acs.chemmater.7b02835}
  {https://doi.org/10.1021/acs.chemmater.7b02835} \BibitemShut {NoStop}%
\bibitem [{\citenamefont {Harashima}\ \emph {et~al.}(2022)\citenamefont
  {Harashima}, \citenamefont {Koga}, \citenamefont {Ni}, \citenamefont
  {Yonehara}, \citenamefont {Katouda}, \citenamefont {Notake}, \citenamefont
  {Matsui}, \citenamefont {Moriya}, \citenamefont {Si}, \citenamefont
  {Hasunuma}, \citenamefont {Uedono},\ and\ \citenamefont
  {Shigeta}}]{Harashima2022}%
  \BibitemOpen
  \bibfield  {author} {\bibinfo {author} {\bibfnamefont {Y.}~\bibnamefont
  {Harashima}}, \bibinfo {author} {\bibfnamefont {H.}~\bibnamefont {Koga}},
  \bibinfo {author} {\bibfnamefont {Z.}~\bibnamefont {Ni}}, \bibinfo {author}
  {\bibfnamefont {T.}~\bibnamefont {Yonehara}}, \bibinfo {author}
  {\bibfnamefont {M.}~\bibnamefont {Katouda}}, \bibinfo {author} {\bibfnamefont
  {A.}~\bibnamefont {Notake}}, \bibinfo {author} {\bibfnamefont
  {H.}~\bibnamefont {Matsui}}, \bibinfo {author} {\bibfnamefont
  {T.}~\bibnamefont {Moriya}}, \bibinfo {author} {\bibfnamefont {M.~K.}\
  \bibnamefont {Si}}, \bibinfo {author} {\bibfnamefont {R.}~\bibnamefont
  {Hasunuma}}, \bibinfo {author} {\bibfnamefont {A.}~\bibnamefont {Uedono}}, \
  and\ \bibinfo {author} {\bibfnamefont {Y.}~\bibnamefont {Shigeta}},\ }in\
  \href {\doibase 10.1109/ISSM55802.2022.10027044} {\emph {\bibinfo {booktitle}
  {2022 International Symposium on Semiconductor Manufacturing (ISSM)}}}\
  (\bibinfo {year} {2022})\ pp.\ \bibinfo {pages} {1--3}\BibitemShut {NoStop}%
\bibitem [{\citenamefont {Wang}\ \emph {et~al.}(2006)\citenamefont {Wang},
  \citenamefont {Zinkevich},\ and\ \citenamefont {Aldinger}}]{Wang2006}%
  \BibitemOpen
  \bibfield  {author} {\bibinfo {author} {\bibfnamefont {C.}~\bibnamefont
  {Wang}}, \bibinfo {author} {\bibfnamefont {M.}~\bibnamefont {Zinkevich}}, \
  and\ \bibinfo {author} {\bibfnamefont {F.}~\bibnamefont {Aldinger}},\ }\href
  {\doibase https://doi.org/10.1111/j.1551-2916.2006.01286.x} {\bibfield
  {journal} {\bibinfo  {journal} {Journal of the American Ceramic Society}\
  }\textbf {\bibinfo {volume} {89}},\ \bibinfo {pages} {3751} (\bibinfo {year}
  {2006})},\ \Eprint
  {http://arxiv.org/abs/https://ceramics.onlinelibrary.wiley.com/doi/pdf/10.1111/j.1551-2916.2006.01286.x}
  {https://ceramics.onlinelibrary.wiley.com/doi/pdf/10.1111/j.1551-2916.2006.01286.x}
  \BibitemShut {NoStop}%
\bibitem [{\citenamefont {Bl\"ochl}(1994)}]{Bloechl1994}%
  \BibitemOpen
  \bibfield  {author} {\bibinfo {author} {\bibfnamefont {P.~E.}\ \bibnamefont
  {Bl\"ochl}},\ }\href {\doibase 10.1103/PhysRevB.50.17953} {\bibfield
  {journal} {\bibinfo  {journal} {Phys. Rev. B}\ }\textbf {\bibinfo {volume}
  {50}},\ \bibinfo {pages} {17953} (\bibinfo {year} {1994})}\BibitemShut
  {NoStop}%
\bibitem [{\citenamefont {Kresse}\ and\ \citenamefont
  {Furthm\"uller}(1996)}]{Kresse1996}%
  \BibitemOpen
  \bibfield  {author} {\bibinfo {author} {\bibfnamefont {G.}~\bibnamefont
  {Kresse}}\ and\ \bibinfo {author} {\bibfnamefont {J.}~\bibnamefont
  {Furthm\"uller}},\ }\href {\doibase 10.1103/PhysRevB.54.11169} {\bibfield
  {journal} {\bibinfo  {journal} {Phys. Rev. B}\ }\textbf {\bibinfo {volume}
  {54}},\ \bibinfo {pages} {11169} (\bibinfo {year} {1996})}\BibitemShut
  {NoStop}%
\bibitem [{\citenamefont {Kresse}\ and\ \citenamefont
  {Joubert}(1999)}]{Kresse1999}%
  \BibitemOpen
  \bibfield  {author} {\bibinfo {author} {\bibfnamefont {G.}~\bibnamefont
  {Kresse}}\ and\ \bibinfo {author} {\bibfnamefont {D.}~\bibnamefont
  {Joubert}},\ }\href {\doibase 10.1103/PhysRevB.59.1758} {\bibfield  {journal}
  {\bibinfo  {journal} {Phys. Rev. B}\ }\textbf {\bibinfo {volume} {59}},\
  \bibinfo {pages} {1758} (\bibinfo {year} {1999})}\BibitemShut {NoStop}%
\bibitem [{\citenamefont {Perdew}\ \emph {et~al.}(1996)\citenamefont {Perdew},
  \citenamefont {Burke},\ and\ \citenamefont {Ernzerhof}}]{Perdew1996}%
  \BibitemOpen
  \bibfield  {author} {\bibinfo {author} {\bibfnamefont {J.~P.}\ \bibnamefont
  {Perdew}}, \bibinfo {author} {\bibfnamefont {K.}~\bibnamefont {Burke}}, \
  and\ \bibinfo {author} {\bibfnamefont {M.}~\bibnamefont {Ernzerhof}},\ }\href
  {\doibase 10.1103/PhysRevLett.77.3865} {\bibfield  {journal} {\bibinfo
  {journal} {Phys. Rev. Lett.}\ }\textbf {\bibinfo {volume} {77}},\ \bibinfo
  {pages} {3865} (\bibinfo {year} {1996})}\BibitemShut {NoStop}%
\bibitem [{\citenamefont {Okhotnikov}\ \emph {et~al.}(2016)\citenamefont
  {Okhotnikov}, \citenamefont {Charpentier},\ and\ \citenamefont
  {Cadars}}]{Okhotnikov2016}%
  \BibitemOpen
  \bibfield  {author} {\bibinfo {author} {\bibfnamefont {K.}~\bibnamefont
  {Okhotnikov}}, \bibinfo {author} {\bibfnamefont {T.}~\bibnamefont
  {Charpentier}}, \ and\ \bibinfo {author} {\bibfnamefont {S.}~\bibnamefont
  {Cadars}},\ }\href@noop {} {\bibfield  {journal} {\bibinfo  {journal} {J.
  Cheminform.}\ }\textbf {\bibinfo {volume} {8}},\ \bibinfo {pages} {17}
  (\bibinfo {year} {2016})}\BibitemShut {NoStop}%
\bibitem [{\citenamefont {Moruzzi}\ \emph {et~al.}(1988)\citenamefont
  {Moruzzi}, \citenamefont {Janak},\ and\ \citenamefont
  {Schwarz}}]{Moruzzi1988}%
  \BibitemOpen
  \bibfield  {author} {\bibinfo {author} {\bibfnamefont {V.~L.}\ \bibnamefont
  {Moruzzi}}, \bibinfo {author} {\bibfnamefont {J.~F.}\ \bibnamefont {Janak}},
  \ and\ \bibinfo {author} {\bibfnamefont {K.}~\bibnamefont {Schwarz}},\ }\href
  {\doibase 10.1103/PhysRevB.37.790} {\bibfield  {journal} {\bibinfo  {journal}
  {Phys. Rev. B}\ }\textbf {\bibinfo {volume} {37}},\ \bibinfo {pages} {790}
  (\bibinfo {year} {1988})}\BibitemShut {NoStop}%
\bibitem [{\citenamefont {Togo}\ and\ \citenamefont {Tanaka}(2015)}]{Togo2015}%
  \BibitemOpen
  \bibfield  {author} {\bibinfo {author} {\bibfnamefont {A.}~\bibnamefont
  {Togo}}\ and\ \bibinfo {author} {\bibfnamefont {I.}~\bibnamefont {Tanaka}},\
  }\href {\doibase https://doi.org/10.1016/j.scriptamat.2015.07.021} {\bibfield
   {journal} {\bibinfo  {journal} {Scripta Materialia}\ }\textbf {\bibinfo
  {volume} {108}},\ \bibinfo {pages} {1} (\bibinfo {year} {2015})}\BibitemShut
  {NoStop}%
\bibitem [{\citenamefont {Tadano}\ \emph {et~al.}(2014)\citenamefont {Tadano},
  \citenamefont {Gohda},\ and\ \citenamefont {Tsuneyuki}}]{Tadano2014}%
  \BibitemOpen
  \bibfield  {author} {\bibinfo {author} {\bibfnamefont {T.}~\bibnamefont
  {Tadano}}, \bibinfo {author} {\bibfnamefont {Y.}~\bibnamefont {Gohda}}, \
  and\ \bibinfo {author} {\bibfnamefont {S.}~\bibnamefont {Tsuneyuki}},\ }\href
  {\doibase 10.1088/0953-8984/26/22/225402} {\bibfield  {journal} {\bibinfo
  {journal} {Journal of Physics: Condensed Matter}\ }\textbf {\bibinfo {volume}
  {26}},\ \bibinfo {pages} {225402} (\bibinfo {year} {2014})}\BibitemShut
  {NoStop}%
\bibitem [{\citenamefont {Kuwabara}\ \emph {et~al.}(2005)\citenamefont
  {Kuwabara}, \citenamefont {Tohei}, \citenamefont {Yamamoto},\ and\
  \citenamefont {Tanaka}}]{Kuwabara2005}%
  \BibitemOpen
  \bibfield  {author} {\bibinfo {author} {\bibfnamefont {A.}~\bibnamefont
  {Kuwabara}}, \bibinfo {author} {\bibfnamefont {T.}~\bibnamefont {Tohei}},
  \bibinfo {author} {\bibfnamefont {T.}~\bibnamefont {Yamamoto}}, \ and\
  \bibinfo {author} {\bibfnamefont {I.}~\bibnamefont {Tanaka}},\ }\href
  {\doibase 10.1103/PhysRevB.71.064301} {\bibfield  {journal} {\bibinfo
  {journal} {Phys. Rev. B}\ }\textbf {\bibinfo {volume} {71}},\ \bibinfo
  {pages} {064301} (\bibinfo {year} {2005})}\BibitemShut {NoStop}%
\bibitem [{\citenamefont {Luo}\ \emph {et~al.}(2009)\citenamefont {Luo},
  \citenamefont {Zhou}, \citenamefont {Ushakov}, \citenamefont {Navrotsky},\
  and\ \citenamefont {Demkov}}]{Luo2009}%
  \BibitemOpen
  \bibfield  {author} {\bibinfo {author} {\bibfnamefont {X.}~\bibnamefont
  {Luo}}, \bibinfo {author} {\bibfnamefont {W.}~\bibnamefont {Zhou}}, \bibinfo
  {author} {\bibfnamefont {S.~V.}\ \bibnamefont {Ushakov}}, \bibinfo {author}
  {\bibfnamefont {A.}~\bibnamefont {Navrotsky}}, \ and\ \bibinfo {author}
  {\bibfnamefont {A.~A.}\ \bibnamefont {Demkov}},\ }\href {\doibase
  10.1103/PhysRevB.80.134119} {\bibfield  {journal} {\bibinfo  {journal} {Phys.
  Rev. B}\ }\textbf {\bibinfo {volume} {80}},\ \bibinfo {pages} {134119}
  (\bibinfo {year} {2009})}\BibitemShut {NoStop}%
\bibitem [{\citenamefont {Miyake}\ \emph {et~al.}(2014)\citenamefont {Miyake},
  \citenamefont {Terakura}, \citenamefont {Harashima}, \citenamefont {Kino},\
  and\ \citenamefont {Ishibashi}}]{Miyake2014}%
  \BibitemOpen
  \bibfield  {author} {\bibinfo {author} {\bibfnamefont {T.}~\bibnamefont
  {Miyake}}, \bibinfo {author} {\bibfnamefont {K.}~\bibnamefont {Terakura}},
  \bibinfo {author} {\bibfnamefont {Y.}~\bibnamefont {Harashima}}, \bibinfo
  {author} {\bibfnamefont {H.}~\bibnamefont {Kino}}, \ and\ \bibinfo {author}
  {\bibfnamefont {S.}~\bibnamefont {Ishibashi}},\ }\href {\doibase
  10.7566/JPSJ.83.043702} {\bibfield  {journal} {\bibinfo  {journal} {J. Phys.
  Soc. Jpn.}\ }\textbf {\bibinfo {volume} {83}},\ \bibinfo {pages} {043702}
  (\bibinfo {year} {2014})}\BibitemShut {NoStop}%
\bibitem [{\citenamefont {Harashima}\ \emph {et~al.}(2016)\citenamefont
  {Harashima}, \citenamefont {Terakura}, \citenamefont {Kino}, \citenamefont
  {Ishibashi},\ and\ \citenamefont {Miyake}}]{Harashima2016}%
  \BibitemOpen
  \bibfield  {author} {\bibinfo {author} {\bibfnamefont {Y.}~\bibnamefont
  {Harashima}}, \bibinfo {author} {\bibfnamefont {K.}~\bibnamefont {Terakura}},
  \bibinfo {author} {\bibfnamefont {H.}~\bibnamefont {Kino}}, \bibinfo {author}
  {\bibfnamefont {S.}~\bibnamefont {Ishibashi}}, \ and\ \bibinfo {author}
  {\bibfnamefont {T.}~\bibnamefont {Miyake}},\ }\href
  {http://scitation.aip.org/content/aip/journal/jap/120/20/10.1063/1.4968798}
  {\bibfield  {journal} {\bibinfo  {journal} {J. Appl. Phys.}\ }\textbf
  {\bibinfo {volume} {120}},\ \bibinfo {eid} {203904} (\bibinfo {year}
  {2016})}\BibitemShut {NoStop}%
\bibitem [{\citenamefont {Harashima}\ \emph {et~al.}(2018)\citenamefont
  {Harashima}, \citenamefont {Fukazawa}, \citenamefont {Kino},\ and\
  \citenamefont {Miyake}}]{Harashima2018}%
  \BibitemOpen
  \bibfield  {author} {\bibinfo {author} {\bibfnamefont {Y.}~\bibnamefont
  {Harashima}}, \bibinfo {author} {\bibfnamefont {T.}~\bibnamefont {Fukazawa}},
  \bibinfo {author} {\bibfnamefont {H.}~\bibnamefont {Kino}}, \ and\ \bibinfo
  {author} {\bibfnamefont {T.}~\bibnamefont {Miyake}},\ }\href
  {https://doi.org/10.1063/1.5050057} {\bibfield  {journal} {\bibinfo
  {journal} {J. Appl. Phys.}\ }\textbf {\bibinfo {volume} {124}},\ \bibinfo
  {pages} {163902} (\bibinfo {year} {2018})}\BibitemShut {NoStop}%
\bibitem [{\citenamefont {Harashima}\ \emph {et~al.}(2020)\citenamefont
  {Harashima}, \citenamefont {Fukazawa},\ and\ \citenamefont
  {Miyake}}]{Harashima2020}%
  \BibitemOpen
  \bibfield  {author} {\bibinfo {author} {\bibfnamefont {Y.}~\bibnamefont
  {Harashima}}, \bibinfo {author} {\bibfnamefont {T.}~\bibnamefont {Fukazawa}},
  \ and\ \bibinfo {author} {\bibfnamefont {T.}~\bibnamefont {Miyake}},\ }\href
  {\doibase https://doi.org/10.1016/j.scriptamat.2020.01.004} {\bibfield
  {journal} {\bibinfo  {journal} {Scripta Materialia}\ }\textbf {\bibinfo
  {volume} {179}},\ \bibinfo {pages} {12 } (\bibinfo {year}
  {2020})}\BibitemShut {NoStop}%
\bibitem [{\citenamefont {Xing}\ \emph {et~al.}(2021)\citenamefont {Xing},
  \citenamefont {Ishikawa}, \citenamefont {Miura}, \citenamefont {Miyake},\
  and\ \citenamefont {Tadano}}]{Xing2021}%
  \BibitemOpen
  \bibfield  {author} {\bibinfo {author} {\bibfnamefont {G.}~\bibnamefont
  {Xing}}, \bibinfo {author} {\bibfnamefont {T.}~\bibnamefont {Ishikawa}},
  \bibinfo {author} {\bibfnamefont {Y.}~\bibnamefont {Miura}}, \bibinfo
  {author} {\bibfnamefont {T.}~\bibnamefont {Miyake}}, \ and\ \bibinfo {author}
  {\bibfnamefont {T.}~\bibnamefont {Tadano}},\ }\href {\doibase
  https://doi.org/10.1016/j.jallcom.2021.159754} {\bibfield  {journal}
  {\bibinfo  {journal} {Journal of Alloys and Compounds}\ }\textbf {\bibinfo
  {volume} {874}},\ \bibinfo {pages} {159754} (\bibinfo {year}
  {2021})}\BibitemShut {NoStop}%
\end{thebibliography}%

\end{document}